\documentclass{article}

\usepackage{arxiv}

\usepackage[utf8]{inputenc} % allow utf-8 input
\usepackage[T1]{fontenc}    % use 8-bit T1 fonts
\usepackage{hyperref}       % hyperlinks
\usepackage{url}            % simple URL typesetting
\usepackage{booktabs}       % professional-quality tables
\usepackage{amsmath,amssymb,amsfonts}
\usepackage{nicefrac}       % compact symbols for 1/2, etc.
\usepackage{microtype}      % microtypography
\usepackage{cleveref}       % smart cross-referencing
\usepackage{graphicx}
\usepackage[numbers,sort&compress]{natbib}
\usepackage{doi}
\usepackage{xcolor}

\title{A Hybrid Classical-Quantum Approach for Multi-Constrained Location Optimization Problem}

\author{
Jorge Saavedra-Benavides \\
Department of Computer Science\\
The Catholic University of America\\
Washington, DC, USA\\
\texttt{saavedrajor@cua.edu}
\And
J. Alejandro Montanez-Barrera \\
Institute for Advanced Simulation\\
Jülich Supercomputing Centre\\
Jülich, Germany\\
\texttt{j.montanez-barrera@fz-juelich.de}
\AND
Alberto Maldonado-Romo \\
T.J. Watson Research Center\\
IBM\\
New York, USA\\
\texttt{Alberto.Maldonado.Romo@ibm.com}
\And
Daniel Sierra-Sosa \\
Department of Computer Science\\
The Catholic University of America\\
Washington, DC, USA\\
\texttt{sierrasosa@cua.edu}
}

\date{}

\renewcommand{\headeright}{Preprint}
\renewcommand{\undertitle}{Preprint}
\renewcommand{\shorttitle}{Hybrid Classical-Quantum MCLP Optimization}

\hypersetup{
  hidelinks,
  pdftitle={A Hybrid Classical-Quantum Approach for Multi-Constrained Location Optimization Problem},
  pdfauthor={Jorge Saavedra-Benavides, J. Alejandro Montanez-Barrera, Alberto Maldonado-Romo, Daniel Sierra-Sosa},
  pdfkeywords={Facility Location Problem, MCLP, QUBO, QAOA, Unbalanced Penalization, Linear Ramp, Warm Starting},
}

\begin{document}
\maketitle

\begin{abstract}
The Maximal Covering Location Problem (MCLP) is an $\mathcal{NP}$-hard Combinatorial Optimization Problem (COP) that aims to determine the optimal facility placements that maximize total coverage. It is characterized by both equality and inequality constraints, which ensure correct coverage but significantly increase the complexity of exploring the solution space as instance size grows. Hybrid quantum-classical approaches might offer a promising alternative to classical optimization methods by enabling the exploration of complex energy landscapes through quantum superposition and probabilistic sampling. 
In this work, the MCLP is formulated as a Quadratic Unconstrained Binary Optimization (QUBO) model, where constraint embedding plays a critical role in solution quality. In particular, Unbalanced Penalization (UP) is employed as an alternative to the Slack Variables (SV) for handling inequality constraints without increasing the number of variables. 
This study focuses on QAOA and one of its variants, the WS-QAOA, which leverages a biased initial state derived from a continuous relaxation of the problem. Additionally, a linear ramp (LR) parameter schedule is incorporated to reduce optimization complexity. The performance of these techniques is evaluated both individually and in combination, as a function of circuit depth $p$ and problem size. Results show that the combined approach of UP, LR, and WS-QAOA consistently improves solution quality and feasibility metrics, while maintaining robust performance as the problem size increases, highlighting its potential within hybrid quantum-classical optimization frameworks.  
%Results show that the combined approach of UP, LR and WS-QAOA consistently improves solution quality and feasibility metrics, while maintaining robust performance as the problem size increases, highlighting its potential for near-term quantum optimization.
\end{abstract}

\keywords{Facility Location Problem \and MCLP \and QUBO \and QAOA \and Unbalanced Penalization \and Linear Ramp \and Warm Starting}

\section{Introduction}
Advances in quantum computing have enabled the development of promising algorithms for tackling optimization problems that are computationally challenging for classical methods. Some studies suggest implementing quantum optimization techniques for problems with direct impact on logistics, such as routing or facility location \cite{gabbassov_transit_2022, osaba_solving_2024, ciacco_quantum_2026}, leveraging superposition and probabilistic sampling to explore complex, non-convex energy landscapes. Facility Location Problems (FLPs) are the class of problems aimed at determining the optimal placement among the available options to place a given public facility, e.g. hospitals or stores~\cite{tavares_maximal_2025, moncayo-martinez_quantum_2026}. Finding optimal solutions of FLPs can significantly improve efficiency and service accessibility. One particular FLP is the Maximal Covering Location Problem (MCLP).

Introduced by Church R. and ReVelle C. \cite{church_maximal_1974}, MCLP is a combinatorial optimization problem (COP) in the $\mathcal{NP}$-hard class \cite{megiddo_maximum_1983, baldomeronaranjo_complexity_2024}. This problem aims at maximizing the total covered demand by selecting a given number of facilities to be located among an eligible set. ReVelle C. explored the linear programming (LP) relaxation of the MCLP. He concluded that the formulation behaves almost like an integer problem even when relaxed~\cite{revelle_facility_1993}. The integer-friendly formulation of the MCLP suggests that, although integrality is not guaranteed, it could be useful as an initial approximation of the solution.

Quantum computing algorithms proposed for solving COP include the Quantum Annealing and Variational Quantum Algorithms (VQA). Within the VQA framework, the two main algorithms are the Variational Quantum Eigensolver (VQE)~\cite{peruzzo_variational_2014, tilly_variational_2022} and the Quantum Approximate Optimization Algorithm (QAOA)~\cite{farhi_quantum_2014}. The foundation of these algorithms is the approximation of the ground state of a cost Hamiltonian using a parametric trial function, or ansatz. The construction of the ansatz depends on each algorithm, while the variational parameters are optimized classically. Among these approaches, QAOA has received significant attention as a result of its potential resilience to noise in near-term quantum devices and its effectiveness in solving COP \cite{bechtold_investigating_2023}.  

A fundamental step in most quantum optimization algorithms consists of mapping the problem cost function, and its constraints, into a Quadratic Unconstrained Binary Optimization (QUBO) form~\cite{glover_quantum_2022}. The most common embedding method for inequality constraints is the Slack Variables (SV) method, but using this method comes at the cost of introducing additional binary variables. Montanez A. \textit{et al.} proposed Unbalanced Penalization (UP). This alternative to SV enables the embedding of inequality constraints in the QUBO formulation without extra binary variables~\cite{montanez-barrera_unbalanced_2024}.  

In recent studies, Giraldo Q.~\textit{et al.} addressed the MCLP employing the SV method for constraint embedding \cite{giraldo-quintero_using_2022}, providing the first empirical evidence of the application of quantum algorithms, such as QA and QAOA, to the MCLP. Their results showed that, under the SV formulation, QAOA exhibited a limited capacity to obtain optimal solutions, while QA achieved more accurate solutions. To overcome the limitations associated with SV, Moncayo L. and He N. \cite{moncayo-martinez_quantum_2026} validated the effectiveness of UP for multi-constrained problems, such as the Minimum Dominating Set problem and the Load Balancing problem. According to their results, UP in combination with QAOA achieved higher feasibility ratios while requiring fewer qubits compared to SV. Another approach explored for a FLP variant, includes a classical preprocessing to reduce the size of the problem \cite{ciacco_quantum_2026}. According to Ciacco A. \textit{et al.} this reduction led to high-quality solutions employing QA. 

The solution quality does not only depend on the problem representation but also on the algorithm. In recent works numerous variants and sub-routines for the QAOA algorithm have been proposed~\cite{blekos_review_2024}. In this context, some of the algorithms proposed are oriented toward reducing the search space by introducing a preserving ansatz, such as the Quantum Alternating Operator Ansatz (QAOA+)~\cite{wang_quantum_2023}. Egger D. \textit{et al.} proposed WS-QAOA that incorporates an ansatz that takes advantage of classical algorithms to generate a good initial state~\cite{egger_warm-starting_2021}. This approach exhibits better solutions than standard QAOA at low depth, which is practically useful when high fidelity quantum resources are limited.

This paper shows how, for a multi-constrained problem such as the MCLP, the use of UP and the integration of classical and quantum subroutines can enhance the performance of QAOA. To the best of our knowledge, no prior work has explored the combined effect of UP formulation and WS-QAOA strategies for the MCLP. In this work, these techniques are further integrated with the Linear Ramp (LR) schedule~\cite{montanez-barrera_towards_2025} to define the variational parameters of QAOA. While each of these approaches has individually demonstrated improvements in solution quality and optimization performance, their combined effect is systematically evaluated in this study.

%Individually, these techniques have demonstrated an improvement on the performance and quality solution of QAOA, and the integration of them is evaluated.

The paper has the following structure. Section~\ref{Sec1} presents the background of this work, including a formal definition of the MCLP, the QUBO formulation, and the QAOA algorithm. Section~\ref{Sec2} describes the methods and the experimental setup. Section~\ref{Sec3} presents the experimental results and their discussion. Finally, section~\ref{Sec4} summarizes the main results and presents the conclusions.

\section{Background} \label{Sec1}

\subsection{MCLP}
Introduced by Church R. and ReVelle C. \cite{church_maximal_1974, church_location_2018}, the Maximal Covering Location Problem (MCLP) is a combinatorial optimization problem that aims to maximize the total covered demand. 

Let $I$ be the set of demand nodes, $J$ the set of eligible facilities, and $P$ the number of facilities to be located. The MCLP can be defined as follows:

\begin{equation} \label{obj}
    \max \sum _{i \in I} a_i y_i,
\end{equation}

where $a_i$ denotes the demand associated with node $i$.

This problem is subject to the following constraints:

\begin{equation} \label{equ}
    \sum _{j \in J} x_j = P,
\end{equation}

and

\begin{equation}
    y_i \leq \sum _{j \in N_i} x_j \quad \forall i \in I,
\end{equation}

where $N_i\subseteq J$ represents the set of facilities that can cover demand node $i$. Furthermore, $y_i$ and $x_j$ are binary decision variables defined as:

\begin{equation}
    y_i\in \{0, 1 \}, \quad \forall i \in I,
\end{equation}

and

\begin{equation}
    x_j\in \{0, 1 \}, \quad \forall j \in J.
\end{equation}

$y_i=1$ if demand node $i$ is covered by at least one selected facility ($x_j=1$ for some $j \in N_i$), and 0 otherwise.

This formulation ensures that exactly $P$ facilities are placed and that a demand node is considered covered only if at least one facility within its coverage set is selected.

\subsection{QUBO/Ising Model}
The QUBO formulation~\cite{lucas_ising_2014} is a widely used representation for expressing COP in terms of binary variables. This formulation can be expressed as a symmetric matrix $\mathbf{Q}\in\mathbb{R}^{n\times n}$, where $n$ is the number of binary variables. The entries $Q_{ij} \in \mathbf{Q}$ encode the information of the problem. Due to the equivalence of minimization and maximization (since $\min{(f_C(\mathbf{x}))} = -\max{(-f_C(\mathbf{x}))}$, where $f_C$ is a cost function), the optimization problem now consists of finding the optimal binary string $\mathbf{x} ^ * \in \{0, 1\}^n$ that minimizes the quadratic form defined by $\mathbf{Q}$:
\begin{equation}
    \begin{split}
    &\underset{\mathbf{x  \in \{0, 1\}^n}}{\min} \; \mathbf{x^T} \mathbf{Q} \mathbf{x} = 
    \\&\underset{\mathbf{x}  \in \{0, 1\}^n}{\min} \left [  \sum _{i=1} ^{n} Q_{ii} x_i +  2 \sum _{i=1} ^{n} \sum _{j>i} ^{n} Q_{ij} x_i x_j \right ].
    \end{split}
\end{equation}

In quantum computing, binary variables are mapped to qubit states, where the computational basis states $|0 \rangle$ and $|1 \rangle$ correspond to eigenstates of the Pauli operator $\sigma ^z$ with eigenvalues $1$ and $-1$, respectively. This formulation can be mapped to the Ising model through the transformation $x_i = \frac{1-z_i}{2}$, where $z_i\in\{-1, 1\}$ are spin variables. This enables the representation of the problem as a Hamiltonian acting on qubits, whose ground state encodes the optimal solution:

\begin{equation} \label{Eq1}
    \hat H_C = \sum _{i=1} ^{n} h_i \sigma _i ^z + \sum _{i=1}^n \sum_{j > i}^n J _{i, j} \sigma _i ^z \sigma _j ^z,
\end{equation}
where $h_i$ and $J_{i,j}$ are the linear and quadratic coefficients that encode the information of the optimization problem.

\subsection{QAOA}
The QAOA~\cite{farhi_quantum_2014} is a VQA designed to approximate the solution of optimization problems encoded in the ground state of a cost Hamiltonian $\hat H_C$, most commonly an Ising Hamiltonian (Eq.~\ref{Eq1}). This algorithm can be interpreted as the discretized approximation of adiabatic evolution~\cite{grant_adiabatic_2020}. The parametrized ansatz of QAOA applies alternating layers of unitary operators to make the system evolve from the ground state of an initial Hamiltonian, also called mixer Hamiltonian, toward the ground state of the cost Hamiltonian. The evolution of the system is limited by the number of layers $p$.  

The mixer Hamiltonian $\hat H_M$ is commonly defined as:
\begin{equation*}
    \hat H_M = -\sum _{i=1}^n \sigma _i ^x,
\end{equation*} 
since its ground state is known and easy to prepare:
\begin{equation}
    |s\rangle = H^{\otimes n} |0 \rangle ^{n} = | + \rangle ^{n},
\end{equation}
where $H$ is the Hadamard gate, and $n$ is the number of qubits needed to represent the solution of the problem.

The parametrized state after applying $p$ alternating layers of unitary operators is:
\begin{equation}
    |\vec \gamma,\vec \beta \rangle = e^{-i\beta_{p-1} \hat H_M} e^{-i\gamma_{p-1} \hat H_C} \cdots e^{-i\beta_0 \hat H_M} e^{-i\gamma_0 \hat H_C} |s \rangle,
\end{equation}
where $\vec \gamma = (\gamma_0, \cdots , \gamma_{p-1})$, and $\vec \beta = (\beta _0 , \cdots , \beta_{p-1})$ are variational parameters.

The expectation value of $H_C$ is estimated through repeated measurements of the parametrized quantum state. In that sense, we are interested in finding the optimal state $| \vec \gamma ^*, \vec \beta ^*\rangle$ that minimizes the expectation value of the cost Hamiltonian,
\begin{equation}
    \langle \vec \gamma, \vec \beta| 
\hat H_C |\vec \gamma,\vec \beta \rangle.
\end{equation}

Variational parameters are optimized using classical optimization techniques, iteratively updating their values to minimize the cost function until the stop criterion is achieved. Since the performance of QAOA is affected by circuit depth and parameter optimization complexity, there is a trade off between these two factors. 

\section{Methods} \label{Sec2}

\subsection{Unbalanced Penalization}
UP is an embedding method for inequality constraints compatible with the QUBO form proposed by Montanez A. \textit{et al.} UP is an alternative to the SV method that does not require extra binary variables. This method is based on the incorporation of an exponential term, $e^{-h(\mathbf{x})}$, into the objective function to penalize infeasible solutions. The function $h(\mathbf{x})$ represents an inequality,
\begin{equation}
    h(\mathbf{x}) = W - \sum _{i=1}^{n} w_i x_i \geq 0,
\end{equation}
where commonly $W, w_i \in \mathbb{R}^+$, and $\mathbf{x} \in \{0, 1 \}^n$.

The penalization mechanism is activated when the constraint is violated, i.e., when $h(\mathbf{x})<0$, in which case $e^{-h(\mathbf{x})}$ increases exponentially. In contrast, for feasible solutions ($h(\mathbf{x}) > 0$), the penalization term decreases exponentially, reducing its impact on the objective function. To obtain a QUBO-compatible formulation, a second-order Taylor expansion is used to approximate the exponential term, assuming small constraint violations:
\begin{equation}
    e^{-h(\mathbf{x})} \approx 1 - h(\mathbf{x}) + \frac{1}{2} h(\mathbf{x}) ^ 2.
\end{equation}

An analogous formulation applies to constraints of the form:
\begin{equation}
    g(\mathbf{x}) = W - \sum _{i=1}^{n} w_i x_i \leq 0,
\end{equation}

where the penalization term would be $e^{g(\mathbf{x})}$.

To enhance flexibility, free parameters $\lambda_1$ and $\lambda_2$ are introduced, allowing the penalization strength to be adjusted. Since constant terms do not affect the minimization of the objective function, the constant in Taylor expansion could be omitted, leading to the following penalization function:
\begin{equation}
    \begin{split}
        &f_{UP}(\lambda_1, \lambda_2, \mathbf{x}) = -\lambda _1 h(\mathbf{x}) + \lambda _2 h(\mathbf{x}) ^ 2 = \\& -\lambda _1 \left( W - \sum _{i=1}^{n} w_i x_i \right) + \lambda _2 \left( W - \sum _{i=1}^{n} w_i x_i \right) ^ 2.
    \end{split}
\end{equation}

As a result, the general expression of the total cost function that incorporates equality and inequality constraints can be expressed as
\begin{equation}\label{tot_cost}
    \begin{split}
    \underset{\mathbf{x  \in \{0, 1\}^n}}{\min} \left[ f_C(\mathbf{x}) + \sum_{i=1}^{l}f_{EQ}^{(i)}(\lambda_0^{(i)}, \mathbf{x}) + \sum_{i=1}^{m}f_{UP}^{(i)}(\lambda_1^{(i)}, \lambda_2^{(i)}, \mathbf{x}) \right],
    \end{split}
\end{equation}

where $f_C(\mathbf{x})$ is the objective function of the problem, e.g. Eq. \ref{obj}; $f_{EQ}(\lambda_0, \mathbf{x})$ is the quadratic penalty for equality constraints. As an example, the penalization term associated with Eq. \ref{equ} can be expressed as:
\begin{equation}
    f_{EQ} = \lambda_0 \left( \sum _{j \in J} x_j - P \right)^2,
\end{equation}
where $\lambda_0$ is a penalization coefficient. The selection of the penalization parameters is crucial to ensure that infeasible solutions are sufficiently penalized relative to the objective function, and remains one of the main challenges in QUBO-based formulations. Additionally, $l$ and $m$ are the number of equality and inequality constraints of the problem. In particular, for the MCLP definition employed in this work, $l=1$ and $m=|J|$. 

\subsection{Linear Ramp}
To complement the constraint encoding strategy, parameter scheduling techniques can be employed to improve the optimization process. Linear Ramp~\cite{montanez-barrera_towards_2025} is a parameter schedule oriented to reduce the dimensionality of the classical optimization problem and has been shown to help mitigate optimization difficulties associated with barren plateaus. The LR schedule is defined in terms of the circuit depth $p$ and two variational parameters $\Delta_\beta$ and $\Delta_\gamma$ that define the structure of the schedule:
\begin{equation}
    \beta_k = \left( 1 - \frac{k}{p} \right) \Delta _{\beta} \quad \text{and} \quad \gamma _k = \left( \frac{k+1}{p} \right) \Delta _\gamma,
\end{equation}

for $k=0, \ldots, p-1$. The reduction in parameter complexity is due to the fact that only $\Delta_\beta$ and $\Delta_\gamma$ must be optimized, instead of the $2 p$ parameters that are optimized without any schedule.

\subsection{Warm Starting}
Among the WS methods that have been proposed for quantum algorithms~\cite{truger_warm-starting_2024}, the WS-QAOA is a biased initial state approach. This variant of QAOA uses a relaxed solution $\mathbf{c}^*\in(0, 1)^n$ obtained classically as an initial state ~\cite{egger_warm-starting_2021}. This solution is incorporated into the initial state

\begin{equation}
    |s\rangle = \bigotimes _{i=1} ^{n} R_y(\theta_i) |0\rangle ^n,
\end{equation}

using the following transformation:

\begin{equation}
    \theta_i =
  \begin{cases}
    2 \arcsin(\sqrt{c_i ^*}) & \text{ if } \epsilon < c_i^* < 1-\epsilon, \\ 
     2 \arcsin(\sqrt{\epsilon}) & \text{ if } c_i^* \leq \epsilon, \\ 
     2 \arcsin(\sqrt{1-\epsilon}) & \text{if } c_i^*\geq 1- \epsilon.
  \end{cases} 
\end{equation}

The regularization parameter $\epsilon\in[0, 0.5]$ provides flexibility in the dynamics of the initial state, particularly when the relaxed solution differs significantly from the optimal binary solution. A mixer Hamiltonian is constructed such that $| s \rangle$ is its ground state, given by:

\begin{equation}
    \hat{H}_M = \sum_{i=1}^n [- \sin(\theta_i) \sigma_i^x - \cos(\theta _i) \sigma_i^z],
\end{equation} 

Then, the corresponding unitary operator associated to $\hat{H}_M$ is defined as: 

\begin{equation}
    \hat{U} _M(\beta_k) = \bigotimes_{i=1}^{n} \left[R_y(\theta_i) R_z(-2 \beta_k) R_y(- \theta_i)\right],
\end{equation}

for $k=0,\ldots,p-1$. 

\subsection{Experimental setup}

\subsubsection{Instance generation}
All instances of MCLP used in this work were systematically generated. Five grid sizes were considered for the MCLP: $2 \times 2$, $2 \times 3$, $2 \times 4$, $3 \times 3$, and $3 \times 4$. The demand associated with each node, $a_i\in[1, 6]$, was sampled from a discrete uniform distribution. For each grid size, 10 independent instances were generated and evaluated. 

A subset of grid nodes was selected as demand points, defining the set $I$, while the set $J$ corresponds to candidate facility locations. Table \ref{tab:T1} contains the cardinality values of the sets $I$ and $J$, the number of facilities to be placed $P$, the number of binary variables $n$, and the rules used to define these values for each grid size.

\begin{table}[!ht]
    \centering
    \caption{Characteristics of the experimental data generated for the MCLP.}
    \begin{tabular}{c|c|c|c|c}
        Grid size \\ ($c \times b$) & $|I|=c\cdot b$ & $ |J| = \left \lfloor \frac{|I|}{2} \right \rfloor$ & $P = \left \lfloor \frac{|J|}{2} \right \rfloor$ & $n = |I| + |J|$ \\ \hline
        $2 \times 2$ & 4 & 2 & 1 & 6 \\
        $2 \times 3$ & 6 & 3 & 1 & 9 \\
        $2 \times 4$ & 8 & 4 & 2 & 12 \\
        $3 \times 3$ & 9 & 4 & 2 & 13 \\
        $3 \times 4$ & 12 & 6 & 3 & 18 
    \end{tabular}
    \label{tab:T1}
\end{table}

\subsubsection{QUBO for MCLP}
For constraint embedding, this work compares the number of qubits required for SV and UP, since previous studies~\cite{montanez-barrera_unbalanced_2024, giraldo-quintero_using_2022} have reported limitations of SV in QAOA performance, partly due to the additional binary variables required (qubits). 

Using the UP method, for simplicity we let $\lambda_1 = \lambda_2$ for all inequality constraints. The randomness of instances of different sizes motivated the hypothesis that penalization coefficients must not be fixed, but depend on the information of the problem. A grid search allowed us to determine the parameters that best adapted to each instance. Thus, the parameters used for the embedding of the constraints, according to Eq. \ref{tot_cost}, are: 
\begin{equation}
    \lambda_0=P \quad \text{and} \quad \lambda_1=\lambda_2=\frac{1}{|I|} \sum_{i=1}^{|I|} a_i.
\end{equation}

\begin{figure}[!ht]
    \centering
    \includegraphics[width=1\linewidth]{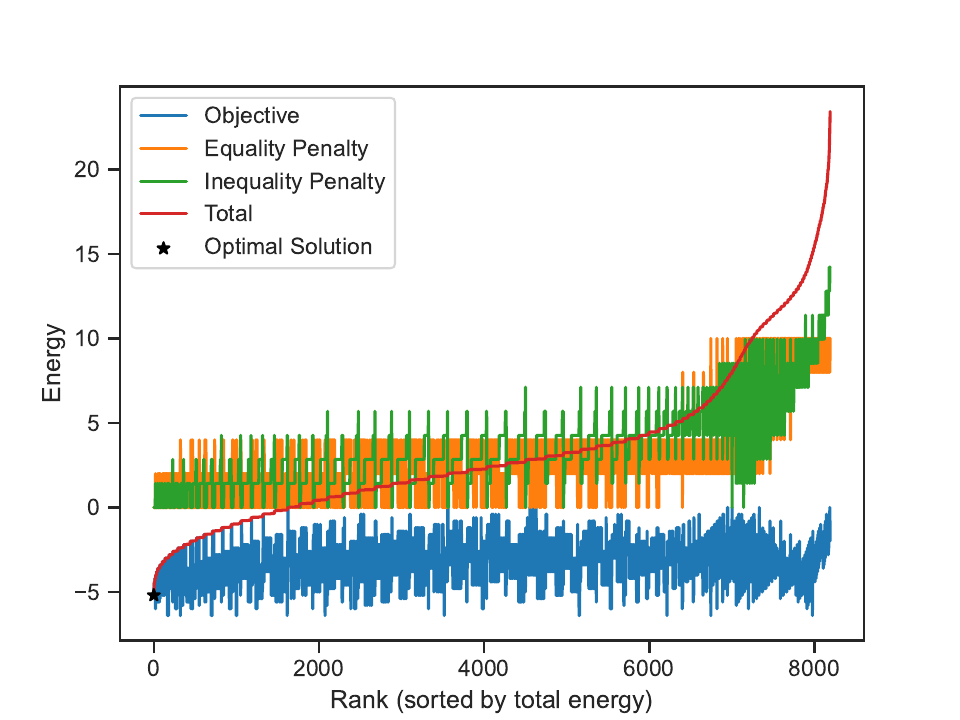}
    \caption{Ranked energy landscape of a $3\times3$ MCLP instance.}
    \label{fig:F4}
\end{figure}

Figure \ref{fig:F4} illustrates the effect of penalization parameters $\lambda_0$, $\lambda_1$, and $\lambda_2$ on the energy landscape of a $3 \times 3$ MCLP instance. The contributions of the objective, the equality constraint penalty, and inequality constraint penalties are shown, along with their combined effect, corresponding to the total cost function. The figure shows that the optimal solution is associated with the lowest energy state, indicating that the selected penalization coefficients effectively enforce feasibility while preserving optimality.

\subsubsection{Experimental workflow}
In this work, we implemented the sub-routines reviewed in Section \ref{Sec2} individually and in combination to assess the impact of each technique on resources required and solution quality. Thus, four experimental models were considered: \textit{UP} (QAOA with UP), \textit{UP\_LR} (QAOA with UP and LR), \textit{UP\_WS} (WS-QAOA with UP), and \textit{UP\_LR\_WS} (WS-QAOA with UP and LR). The regularization parameter used in the models that incorporate WS-QAOA (\textit{UP\_WS} and \textit{UP\_LR\_WS}) was $\epsilon=0.001$. 

\textsc{COBYLA} was used to optimize the variational parameters in all configurations. The stopping criterion for the classical optimizer was $1,000$ iterations with a tolerance of $10^{-4}$. \textsc{CPLEX}~\cite{noauthor_ibm_2025} was used to obtain both the relaxed solution for WS-QAOA and the optimal solution that served as a reference. 

\begin{figure}[!hb]
    \centering
    \includegraphics[width=0.78\linewidth]{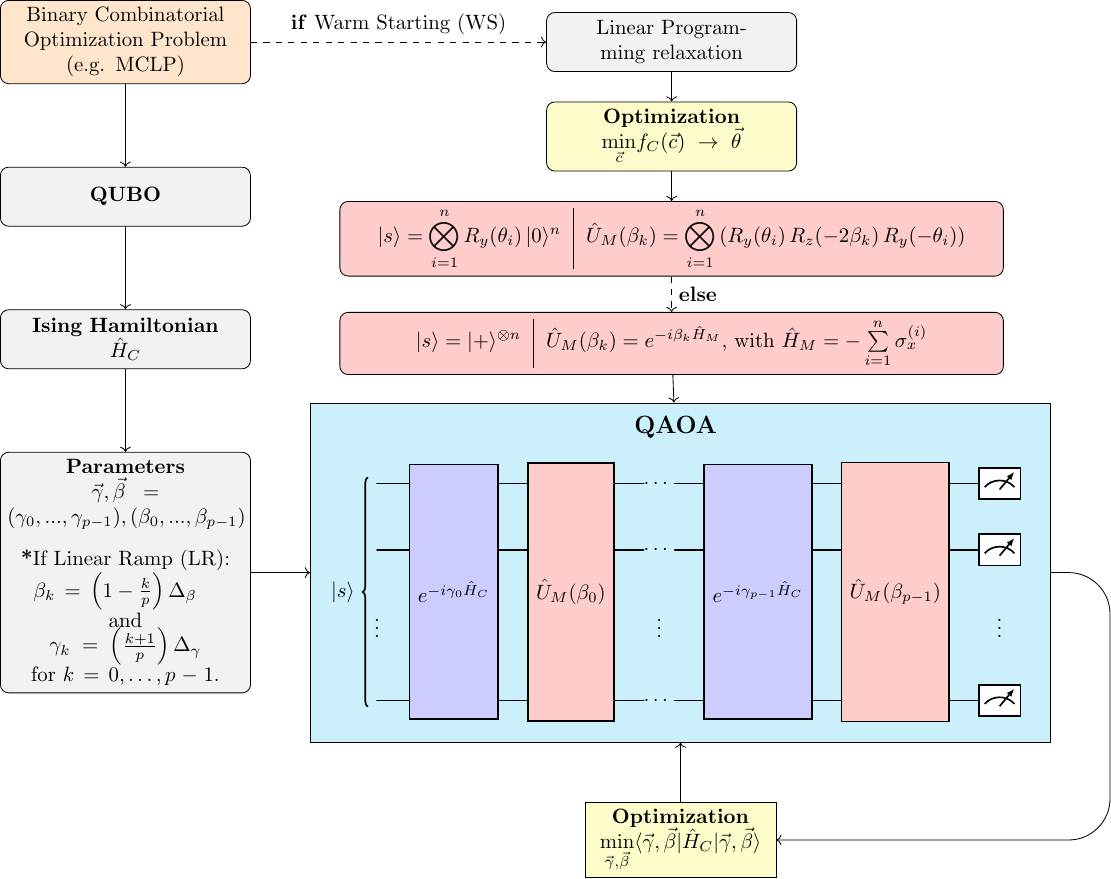}
    \caption{Experimental workflow of the integration WS-QAOA and QAOA with the proposed integrations.}
    \label{fig:Workflow}
\end{figure}

Figure \ref{fig:Workflow} summarizes the experimental workflow followed in this work. 

\subsubsection{Evaluation metrics}
We evaluated the experimental models under the following performance metrics:

\begin{itemize}
    \item{Approximation ratio $r$:}

    \begin{equation}
        r(\psi) = \frac{\langle \psi | \hat H_C | \psi \rangle - e_{\text{max}}}{e_{\text{min}} - e_{\text{max}}},
    \end{equation}

    where $e_{\text{min}}$ and $e_{\text{max}}$ denote the minimum and maximum eigenvalues of the cost Hamiltonian, respectively.
    
    \item{Probability of obtaining the optimal solution $P(\mathbf{x}^*)$:}
    \begin{equation}
        P(\mathbf{x}^*) = | \langle \psi | \mathbf{x}^* \rangle | ^2,
    \end{equation}

    where $\mathbf{x}^* \in \{0, 1 \}^n$ denotes the optimal solution.

    \item{Feasibility ratio $FR$:}

    \begin{equation}
        FR(\mathbf{\psi}) = \frac{\text{Number of feasible samples}}{\text{Total samples}}.
    \end{equation}

    \item{Number of function evaluations performed by the classical optimizer.}
\end{itemize}

It is important to note that the approximation ratio is computed with respect to the penalized cost Hamiltonian. Therefore, its value depends on the selection of the penalization coefficients, which influence the relative weight between feasibility and optimality. In particular, high approximation ratios do not necessarily imply high-quality feasible solutions if the penalization terms dominate the objective function. For this reason, the following section presents the results and interpretation of these metrics jointly to assess the quality of the solution obtained by the different models.

\section{Results and discussion} \label{Sec3}
% Please use ``soft'' (e.g., \verb|\eqref{Eq}|) cross references

\subsection{SV vs UP}
\begin{figure}[!hb]
    \centering
    \includegraphics[width=0.88\linewidth]{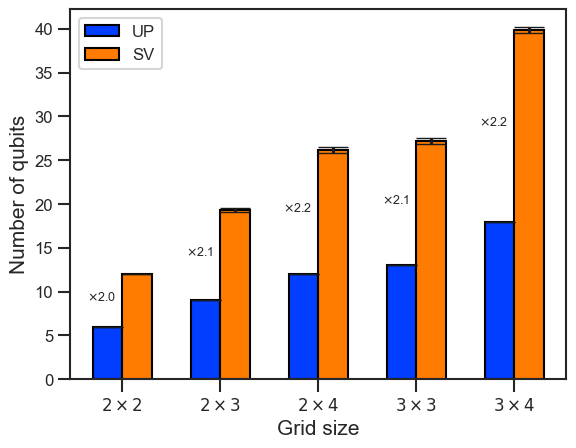}
    \caption{Comparison of the number of qubits required for SV and UP across MCLP instances with different grid sizes.}
    \label{fig:F0}
\end{figure}

Figure~\ref{fig:F0} compares the number of qubits required for the UP and SV methods for each grid size. For the instances considered in this study, the SV method requires approximately twice as many variables as UP, resulting in higher qubit requirements. 

As reported in~\cite{montanez-barrera_unbalanced_2024, moncayo-martinez_quantum_2026}, the additional variables introduced by the SV method increase the size of the search space and the associated computational cost. This makes UP a more suitable approach for embedding inequality constraints into the QUBO formulation, particularly in the context of current non-fault-tolerant quantum hardware, where resource limitations are critical.

\subsection{Effect of circuit depth \texorpdfstring{$p$}{p}}
\begin{figure}[!b]
    \centering
    \includegraphics[width=1\linewidth]{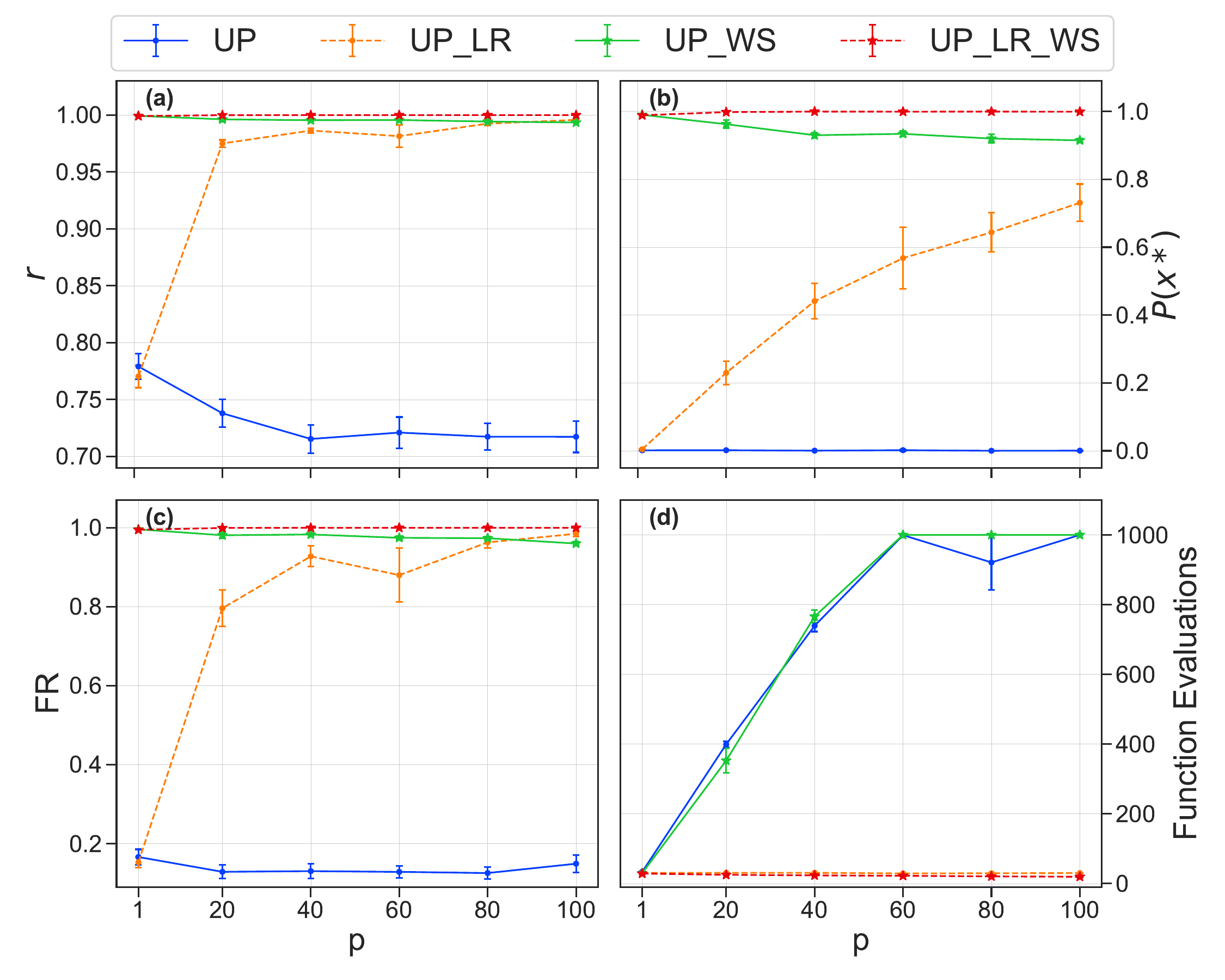}
    \caption{Average values and standard deviation of \textbf{(a)} the approximation ratio $r$, \textbf{(b)} the probability of obtaining the optimal solution $P(\mathbf{x^*})$, \textbf{(c)} the feasibility ratio $FR$, \textbf{(d)} and the number of function evaluations as a function of $p$, for 10 randomly generated instances of the MCLP with a grid size of $2 \times 4$ (12 qubits).}
    \label{fig:F2}
\end{figure}

\begin{figure}[!ht]
    \centering
    \includegraphics[width=1\linewidth]{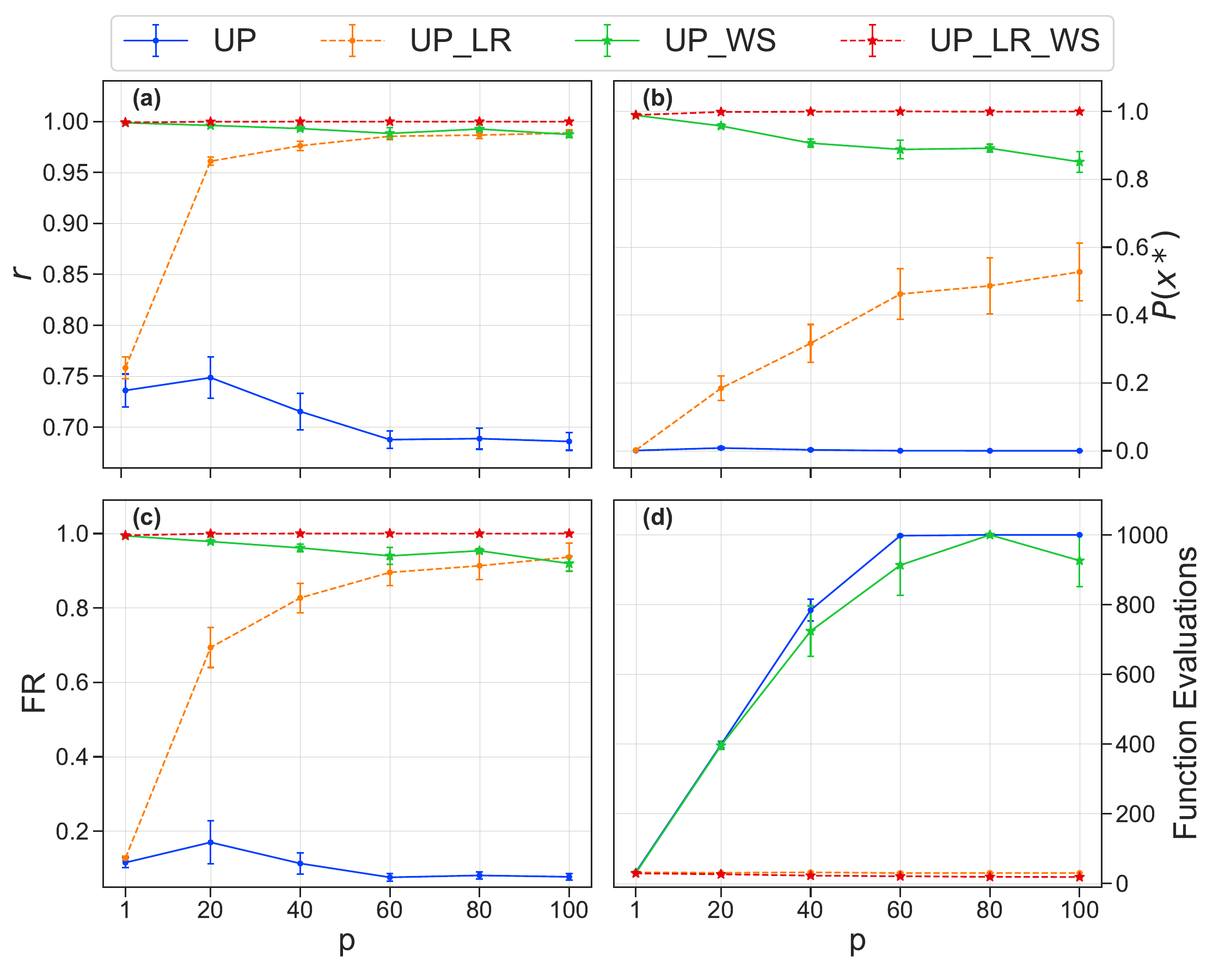}
    \caption{Average values and standard deviation of \textbf{(a)} the approximation ratio $r$, \textbf{(b)} the probability of obtaining the optimal solution $P(\mathbf{x^*})$, \textbf{(c)} the feasibility ratio $FR$, \textbf{(d)} and the number of function evaluations as a function of $p$, for 10 randomly generated instances of the MCLP with a grid size of $3 \times 3$ (13 qubits).}
    \label{fig:F3}
\end{figure}

Figure~\ref{fig:F2} and Figure~\ref{fig:F3} compare the performance of the different models evaluated in this work, for grid sizes $2 \times 4$ and $3 \times 3$, respectively. In both figures, the model \textit{UP\_LR\_WS} obtained the best results in all metrics evaluated, achieving average values of $r$, $P(\mathbf{x}^*)$, and $FR$ close to optimal even for a small $p$. The good performance of the models that incorporated WS-QAOA can be attributed to the high-quality initial state obtained classically from continuous relaxation of the problem. In contrast with \textit{UP\_LR\_WS}, the model \textit{UP\_WS} also presented average values of $r$ and $FR$ above $0.90$ for the grid size $2 \times 4$ and above $0.80$ for the grid size $3 \times 3$; however, its performance degraded as $p$ increased. 

%Correct

Similarly to \textit{UP\_LR}, \textit{UP\_LR\_WS} required fewer function evaluations compared to the models that do not incorporate LR. This result is attributed to the optimization complexity associated with the number of parameters, since LR requires a constant number of parameters and, without LR this number increases linearly with $p$.

The model \textit{UP} exhibits the worst results in the evaluation metrics, since it only incorporates the standard version of QAOA. Furthermore, the model \textit{UP\_LR} supports the effectiveness of the LR schedule as a subroutine to set the variational parameters for QAOA. This model achieves average values of $r$ and $FR$ near 1.0, and the probabilities of obtaining the optimal solution around $70 \%$ with $p=100$.

\subsection{Effect of problem size}
\begin{figure}[!ht]
    \centering
    \includegraphics[width=1\linewidth]{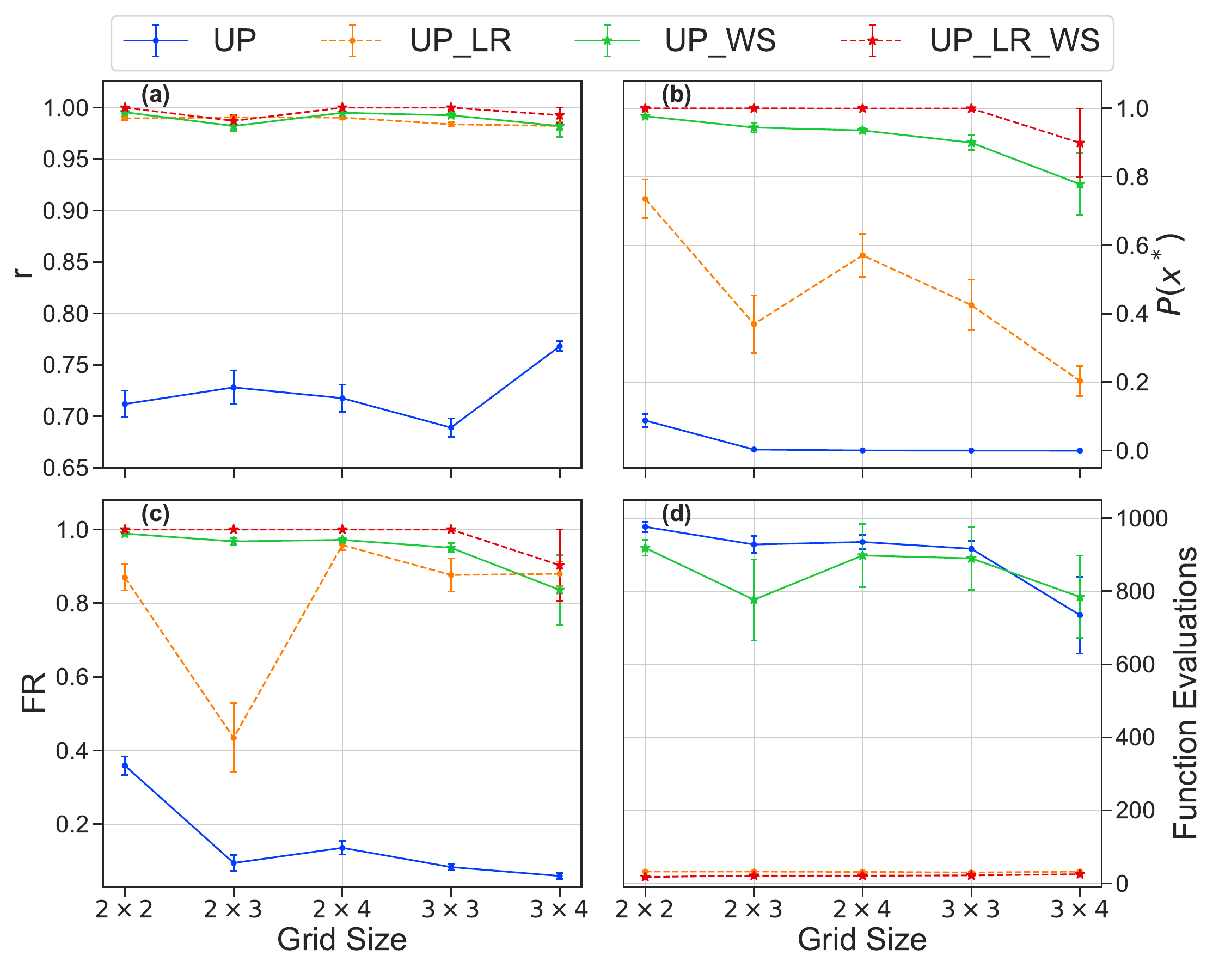}
    \caption{Average values and standard deviation of \textbf{(a)} the approximation ratio $r$, \textbf{(b)} the probability of obtaining the optimal solution $P(\mathbf{x^*})$, \textbf{(c)} the feasibility ratio $FR$, \textbf{(d)} and the number of function evaluations as a function of problem size, for 10 randomly generated instances of the MCLP with $p=50$.}
    \label{fig:F1}
\end{figure}

Figure~\ref{fig:F1} shows the average results obtained by the models across different performance metrics as a function of the size of the instances. All models were evaluated with $p=50$ to ensure equal conditions. In agreement with the results presented in Figure~\ref{fig:F2} and Figure~\ref{fig:F3}, \textit{UP\_LR\_WS} is the model that shows the best performance in the metrics evaluated.

Figure~\ref{fig:F1}.a shows that the approximation ratio $r$ values of the model \textit{UP\_LR} remained above $0.97$ despite the reduction of $P(\mathbf{x^*})$, as the problem size increases. This indicates that LR, in combination with UP, enables the algorithm to approximate low-energy states effectively, even when the probability of sampling the optimal solution decreases.

%In particular, Figure \ref{fig:F1}.b shows how the different models decrease their ability to obtain the optimal solution as the instances increase in size.

In particular, Figure \ref{fig:F1}.b presents the average probabilities of obtaining the optimal solution for the models in the evaluated instances, showing a general tendency to reduce this capacity as the problems increase in size. Nevertheless, the combined approach \textit{UP\_LR\_WS} exhibits more stable performance in obtaining the optimal solution across the evaluated instances. Furthermore, the models that incorporated WS-QAOA maintained a higher probability of obtaining the optimal solution in all instances; nevertheless, these models presented a reduction in $P(\mathbf{x^*})$ (Figure~\ref{fig:F1}.b)  and $FR$ (Figure~\ref{fig:F1}.c) of $10\%$ in $3 \times 4$ instances that is attributed to the quality of the initial state. This highlights the strong dependence of WS-QAOA on the quality of the relaxed solution used as a bias for the initial state, as reported in recent studies~\cite{okada_systematic_2024}.

Figure~\ref{fig:F1}.d shows that the number of function evaluations was significantly higher in models that do not incorporate LR. This number remained practically constant, since the classical optimization complexity depends on the number of parameters.

\section{Conclusions} \label{Sec4}
This work demonstrates how the integration of multiple hybrid quantum-classical approaches can be applied to address combinatorial optimization problems. In particular, we study the performance of UP in a multi-constrained problem, the MCLP. We integrate UP within QAOA and its variant WS-QAOA, and incorporate LR schedule to reduce the optimization complexity of the variational parameters. 

The experimental framework proposed in this work contributes to the study of hybrid quantum-classical techniques for addressing $\mathcal{NP}$-hard problems. The results demonstrate that the integration of these three techniques, UP, LR and WS-QAOA, achieves high probabilities of obtaining the optimal solution while maintaining approximation of low-energy states, for this particular problem. Additionally, the UP method reduces the number of required qubits compared to the Slack Variables approach, while LR enables efficient optimization with reduced parameter complexity. These results indicate that improved encoding (UP), informed initialization (WS), and reduced parameterization (LR) provide complementary advantages for enhancing QAOA performance.

Nevertheless, the main limitation of WS-QAOA is the high dependence on the quality of the relaxed solution used as a bias. Although this work proposes a heuristic rule for the penalization coefficients based on the problem structure, this remains a limitation in many QUBO formulations that open the possibility of further studies.

% number, as in \cite{b3}---do not use ``Ref. \cite{b3}'' or ``reference \cite{b3}'' except at 
% the beginning of a sentence: ``Reference \cite{b3} was the first $\ldots$''

\bibliographystyle{unsrtnat}
\bibliography{references}

\end{document}